\documentclass[12pt]{article}
\def\ypiz_2gg{$\pi^0 \rightarrow \gamma\gamma$}
\newcommand {\charex} {$K^+\mathrm{Xe} \rightarrow K^0 p \mathrm{Xe}'$}
\newcommand {\under} {$K^+n \rightarrow K^0 p$}
\newcommand {\dimass} {$m(pK^0_S)$}

\newcommand {\mtar} {$m^\mathrm{eff}_\mathrm{targ}$}

\begin {document}

\title
{Formation of a narrow baryon resonance with
positive strangeness in $K^+$ collisions with Xe nuclei}

\author{
DIANA Collaboration\\
V.V. Barmin$^a$, 
A.E. Asratyan$^a$,
V.S. Borisov$^a$, 
C. Curceanu$^b$, \\
G.V. Davidenko$^a$, 
A.G. Dolgolenko$^{a,}$\thanks{Corresponding author. E-mail address:
dolgolenko@itep.ru.},
C. Guaraldo$^b$, 
M.A. Kubantsev$^a$, \\
I.F. Larin$^a$, 
V.A. Matveev$^a$, 
V.A. Shebanov$^a$, 
N.N. Shishov$^a$, \\
L.I. Sokolov$^a$,
G.K. Tumanov$^a$,
and V.S. Verebryusov$^a$ \\
\normalsize {$^a$ \it Institute of Theoretical and Experimental Physics,
Moscow 117259, Russia}\\
\normalsize {$^b$ \it Laboratori Nazionali di Frascati dell' INFN,
C.P. 13-I-00044 Frascati, Italy}
}                                          
\date {\today}
\maketitle

\begin{abstract}
The data on the charge-exchange reaction \charex, obtained with the 
bubble chamber DIANA, are reanalyzed using increased statistics and updated 
selections. Our previous evidence for formation of a narrow $pK^0$ 
resonance with mass near 1538 MeV is confirmed and reinforced.
The statistical significance of the signal reaches some 8$\sigma$ 
(6$\sigma$) when estimated as $S /\sqrt{B}$ ($S /\sqrt{B+S}$). The
mass and intrinsic width of the $\Theta^+$ baryon are measured as
$m = 1538\pm2$ MeV and $\Gamma = 0.39\pm0.10$ MeV.
\end{abstract}

\newpage

     The baryons built of four quarks and an antiquark as the lowest 
Fock component, referred to as pentaquarks, are not forbidden by theory 
and have been discussed for many years \cite{forerunners}. Some properties 
of light pentaquark baryons forming the lowest $SU(3)$ multiplet, the
antidecuplet, have been predicted by Diakonov, Petrov, and Polyakov
in the framework of the chiral quark--soliton model \cite{DPP}. In 
particular, they predicted $m \simeq 1530$ MeV and $\Gamma < 15$ MeV 
for the mass and width of the explicitly exotic baryon with positive
strangeness, the $\Theta^+(uudd\bar{s})$ that should decay to the $nK^+$
and $pK^0$ final states. More recent theoretical 
analyses suggest that the $\Theta^+$ intrinsic width may be as small
as $\sim 1$ MeV or even less \cite{width}. Narrow peaks near 1540 MeV 
in the $nK^+$ and $pK^0$ mass spectra were initially detected in
low-energy photoproduction \cite{Nakano-old} and in the charge-exchange
reaction $K^+n \rightarrow pK^0$ \cite{DIANA-old}. Subsequently, both
experiments were able to confirm their initial observations
\cite{Nakano,DIANA}. Moreover, increasing the statistics of the 
charge-exchange reaction allowed DIANA to directly estimate
the $\Theta^+$ intrinsic width: $\Gamma = 0.36\pm0.11$ MeV \cite{DIANA}.
Other searches for the $\Theta^+$ baryon in different reactions and
experimental conditions yielded both positive and negative results,
see the review papers \cite{Burkert} and \cite{Danilov-Mizuk} and
references therein. The bulk of null results can be probably explained
by the extreme smallness of the $\Theta^+$ width that implies the 
smallness of production cross-sections \cite{Diakonov}. 

     The charge-exchange reaction \under\ on a bound neutron, that 
is investigated by DIANA and BELLE \cite{BELLE}, is particularly 
interesting because it allows to probe the $\Theta^+$ intrinsic width 
in a model-independent manner. The existing data on low-energy $K^+D$
scattering have been found to leave room for a $pK^0$ resonance with
mass near 1540 MeV, provided that its width is less than 1 MeV
\cite{Strakovsky,Cahn-Trilling,Sibirtsev-width,Gibbs,Azimov}.
An important advantage of the reaction \under\ is that the 
strangeness of the final-state $pK^0_S$ system is {\it a priori} known 
to be positive. In this paper, the DIANA data on the charge-exchange 
reaction \charex\ are reanalyzed using increased statistics and updated 
selections. 

     The detector and the experimental procedure have been detailed
in \cite{DIANA} and references therein. Briefly, the non-magnetic
bubble chamber DIANA was filled with liquid Xenon and exposed to a 
separated $K^+$ beam with $p = 850$ MeV from the 10-GeV proton 
synchrotron at ITEP, Moscow. A $K^+$ traveling through the chamber
volume is continuously decelerated by ionization, and its momentum at
collision point is a tabulated function of its path in liquid
Xenon. Thereby, we are able to scan $K^+$Xe collisions over an extended
interval of $K^+$ momentum between zero and some 700 MeV. On the other
hand, soft charged secondaries are efficiently identified by bubble
sensity and momentum-analyzed by range in Xenon. (Note that this simple
and elegant scheme will be imitated in future high-statistics studies
of the charge-exchange reaction, see \cite{Muramatsu}.) The estimate
of $K^+$ momentum based on measured position of the interaction vertex
has been verified by detecting and reconstructing the 
$K^+ \rightarrow \pi^+\pi^+\pi^-$ decays in flight. 

     Scanning of the film yielded  nearly 55 000 events 
with visible $K^0_S$ decays, $K^0_S \rightarrow \pi^+\pi^-$ and 
$K^0_S \rightarrow \pi^0\pi^0$, in the full fiducial volume of the
bubble chamber. These $K^0_S$ decays could be associated with primary 
$K^+$Xe vertices with various multiplicities of secondary particles. 
Finally, events with a single proton and $K^0_S \rightarrow \pi^+\pi^-$ 
in the final state were selected as candidates for the charge-exchange 
reaction \under\ on a bound neutron. The selected events are then fully 
measured and reconstructed in space. These measurements are still
in progress. The quality of the data 
is best reflected by experimental resolution on effective mass of the 
$pK^0$ system, estimated as $\sigma_m \simeq 3.5$ MeV by error
propagation for observed events.

\begin{figure}[h]

\vspace{6cm}
\includegraphics{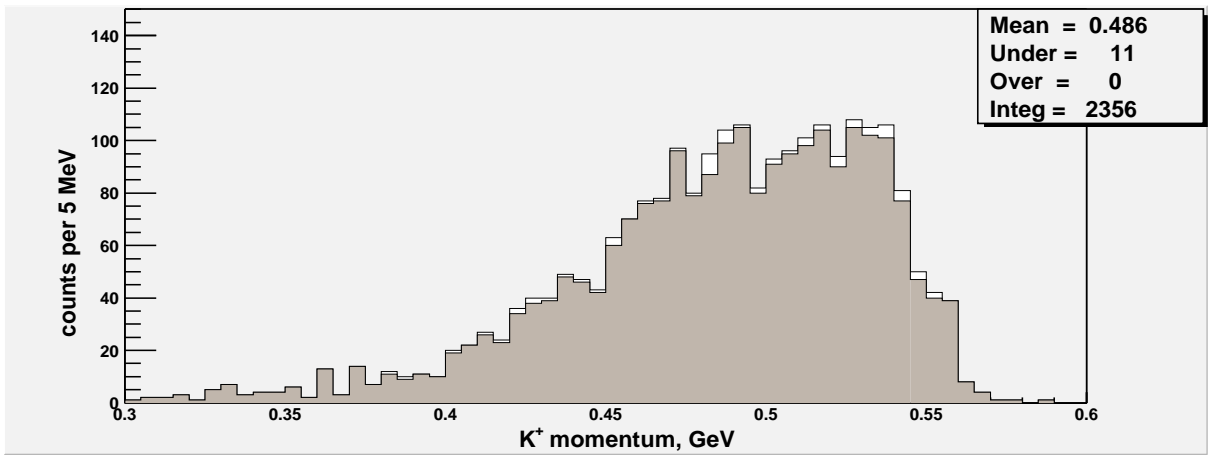}
\caption
{The incident $K^+$ momentum for measured events of the reaction \charex.
The shaded histogram results from restricting the $K^0_S$ proper lifetime:
$\tau < 3\tau_0$.}
\label{pbeam}
\end{figure}
     At present, measurements have been restricted to the region
$L_K > 520$ mm, where $L_K$ is the length of the $K^+$ path in liquid
Xenon before the collision. (Note that there is no one-to-one 
correspondence between $L_K$ and $K^+$ momentum, because the original
beam momentum varied by some $\pm20$ MeV in different exposures.)
Shown in Fig.~\ref{pbeam} is the lab momentum of the incident
$K^+$ for all measured events with $K^0_S$ and proton momenta above 160
and 170 MeV, respectively (instrumental thresholds). Note that the 
statistics of the charge-exchange reaction has been increased by some 
11\% as compared to \cite{DIANA}, and that newly measured events are 
mostly in the high-momentum region.
For further rejection of $K^0_S$ mesons that may have scattered by small
angles in liquid Xenon but passed the pointback criteria, we then apply
the selection
$\tau < 3\tau_0$
where $\tau$ is the $K^0_S$ measured proper lifetime and $\tau_0$ is its 
tabulated mean value. 

\begin{figure}[!t]

\vspace{14cm}
\includegraphics{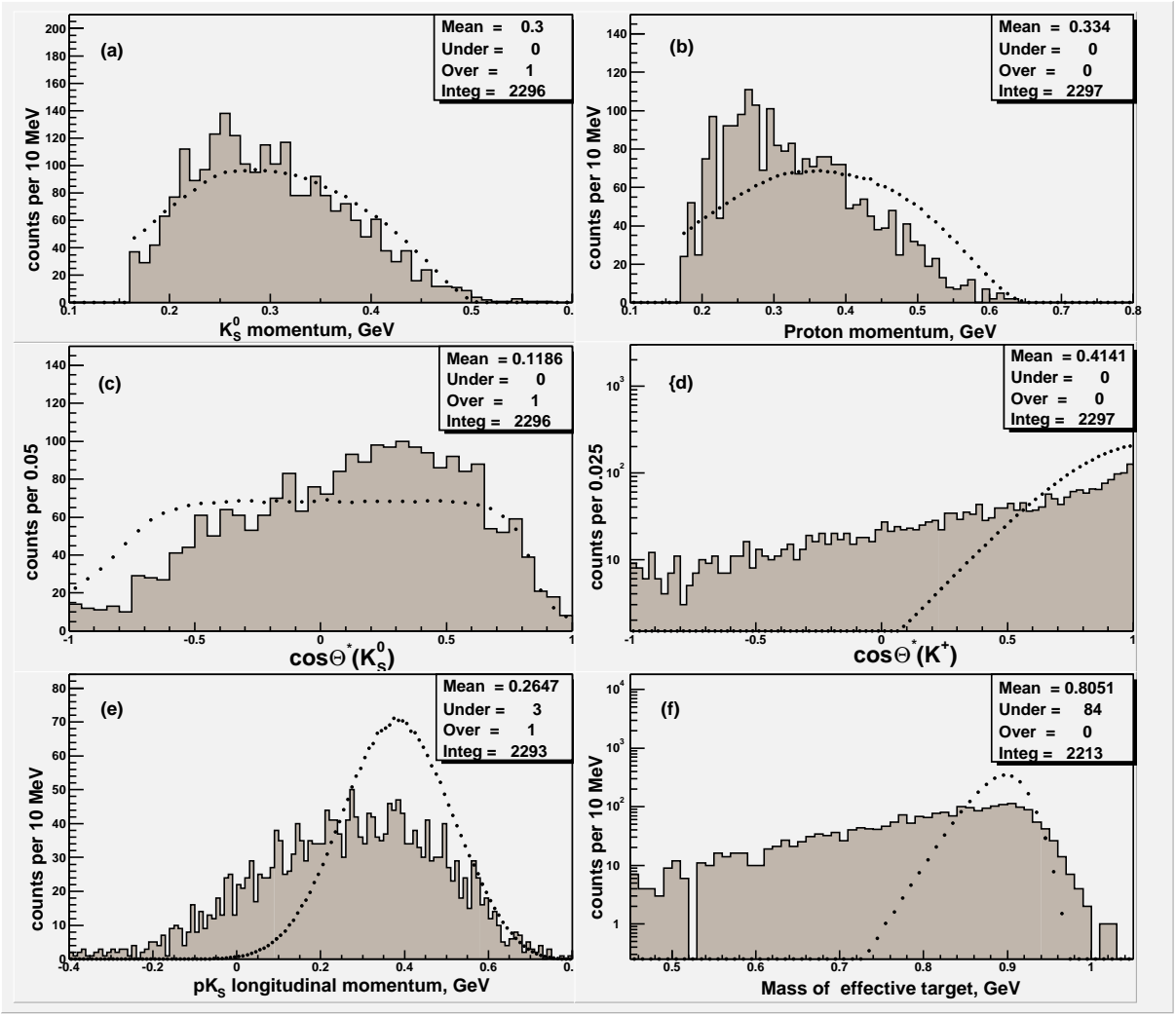}
\caption
{Laboratory momenta of the $K^0_S$ (a) and proton (b),
the cosines of the $K^0_S$ angle (c) and of the $K^+$ angle (d)
in the $pK^0_S$ rest frame, the longitudinal momentum of the 
$pK^0_S$ system in the lab frame (e), and the effective target 
mass \mtar\ (f). The dotted histograms are the 
corresponding distributions of simulated events prior to any proton 
or $K^0_S$ rescattering in nuclear matter, that have been normalized 
to the number of measured events.}
\label{momenta}
\end{figure}

     Figure \ref{momenta} shows the distributions of measured $pK^0_S$
events with $\tau < 3\tau_0$ in the following kinematic variables :
\begin{itemize}
\item
laboratory momenta of the $K^0_S$ and proton ;
\item
in the $pK^0_S$ rest frame, the $K^0_S$ and $K^+$ angles with respect
to the $pK^0_S$ boost from the laboratory frame, denoted as
$\Theta^*(K^0_S)$ and 
$\Theta^*(K^+)$ ;
\item
in the laboratory frame, the longitudinal momentum of the $pK^0_S$ 
system as a whole, $p_L(pK^0_S)$ ;
\item
the effective target mass \mtar, computed assuming binary kinematics.
\end{itemize}
Also shown are corresponding distributions of simulated events prior 
to any proton or $K^0_S$ rescattering in 
nuclear matter,\footnote{The non-resonant 
charge-exchange reaction \under\ in nuclear 
environment is simulated using a simple
Monte-Carlo procedure. The total cross-section
as a function of collision energy is parametrized
using the existing data \cite{cross-section},
and the angular distribution in the $pK^0_S$ rest
frame is assumed to be isotropic. The
total energy of a bound neutron is parametrized 
as $E_n = m_n - \epsilon$ where $m_n$ is the mass of 
a free neutron and $\epsilon\simeq 22$ MeV is the mean 
binding energy for nucleons in the Xenon nucleus.
For the same nucleus, we 
use a realistic form of the Fermi-momentum
distribution with maximum near 160 MeV/c. 
Pauli blocking for secondary protons is approximated
by the cut $p_p > 170$ MeV on proton momentum.
The flux of incident $K^+$ mesons as a function of 
$K^+$ momentum at collision point is inferred
from the observed distribution of $K^+$ range 
in Xenon before interaction or decay, see 
\cite{DIANA-old}. The experimental uncertainties
and measurements errors are included in the 
simulation. Rescattering in the nucleus 
is not accounted for.}
that have been normalized to the number of measured events. 
The $K^0_S$
and proton momenta are seen to be reduced by rescattering, see 
Figs.~\ref{momenta}a and \ref{momenta}b. The effect
is stronger for protons in agreement with the assumption that the
free path in nuclear matter is bigger for kaons than for protons
\cite{Sibirtsev}. Likewise, the $K^0_S$ tends to travel in the forward 
hemisphere in the $pK^0_S$ frame, see Fig.~\ref{momenta}c, because 
rescattering in nuclear matter is more probable for protons than for 
kaons. The $K^+$ travels exclusively in the forward direction
$\cos\Theta^*(K^+) > 0$ for simulated events, see Fig.~\ref{momenta}d,
but also in the backward hemisphere for live events. 
For the $pK^0_S$ longitudinal momentum plotted in 
Fig.~\ref{momenta}e, the simulation predicts that unrescattered 
events should populate the region $p_L > 0$, whereas the 
live events extend down to some -300 MeV. The distribution of the
effective target mass \mtar\ is much broader for the live data than
for simulated events, see Fig.~\ref{momenta}f, because rescattering
disrupts the two-body kinematics of the reaction \under.

\begin{figure}[!h]

\vspace{5.5cm}
\includegraphics{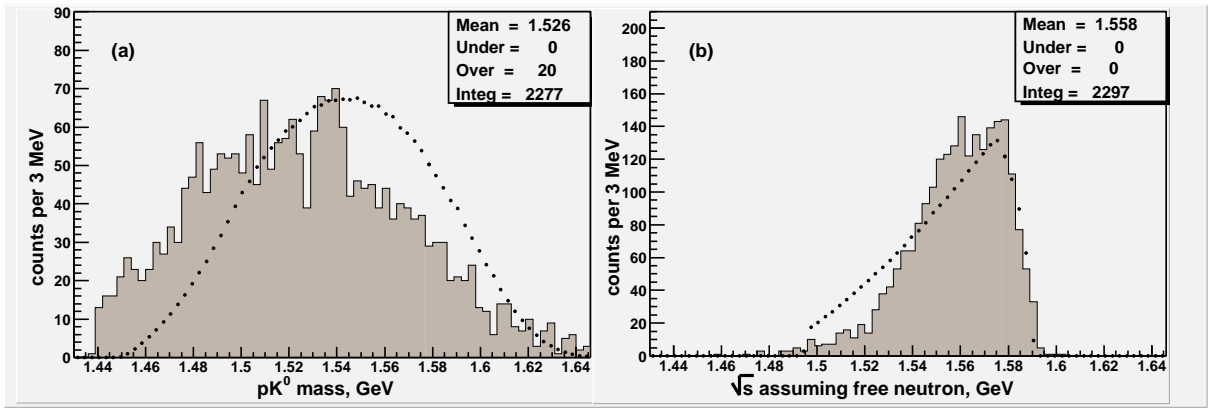}
\caption
{The effective mass of the detected $pK^0_S$ system (a) and the 
center-of-mass energy of the $K^+n$ collision assuming a stationary 
free neutron as the target (b). The dotted histograms depict the
distributions of simulated unrescattered events normalized to the
number of measured events.}
\label{dimass-and-s}
\end{figure}

     Compared in Fig.~\ref{dimass-and-s} are the distributions of two
interrelated variables: the effective mass of the detected $pK^0_S$
system, \dimass, and the center-of-mass energy of the $K^+n$ collision 
assuming a stationary free neutron as the target, $\sqrt{s}$. That the 
\dimass\ spectrum is broader than the $\sqrt{s}$ distribution is due 
to Fermi motion of the target neutron, as well as to rescattering of the 
secondary proton and $K^0$ in nuclear medium that distorts the mass of 
the originally formed $pK^0_S$ system. 

\begin{figure}[h]

\vspace{10cm}
\includegraphics{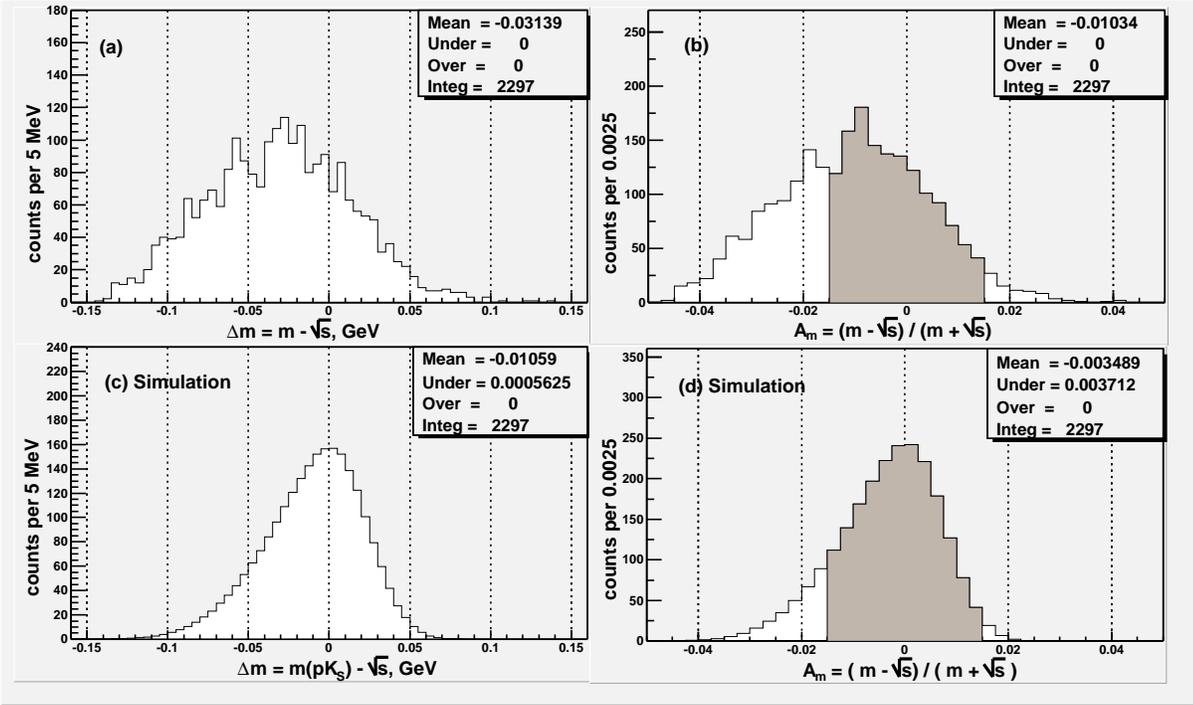}
\caption
{The mass difference $\Delta m = m - \sqrt{s}$ (a) and the asymmetry
$A_m = ( m - \sqrt{s} ) / ( m + \sqrt{s} )$ (b) for observed events.
Corresponding distributions of simulated events
are shown in (c) and (d). Shaded areas in (b) and (d) are for
$|A_m| < 0.015$.}
\label{difma-and-asymma}
\end{figure}

\begin{figure}[!h]

\vspace{5.5cm}
\includegraphics{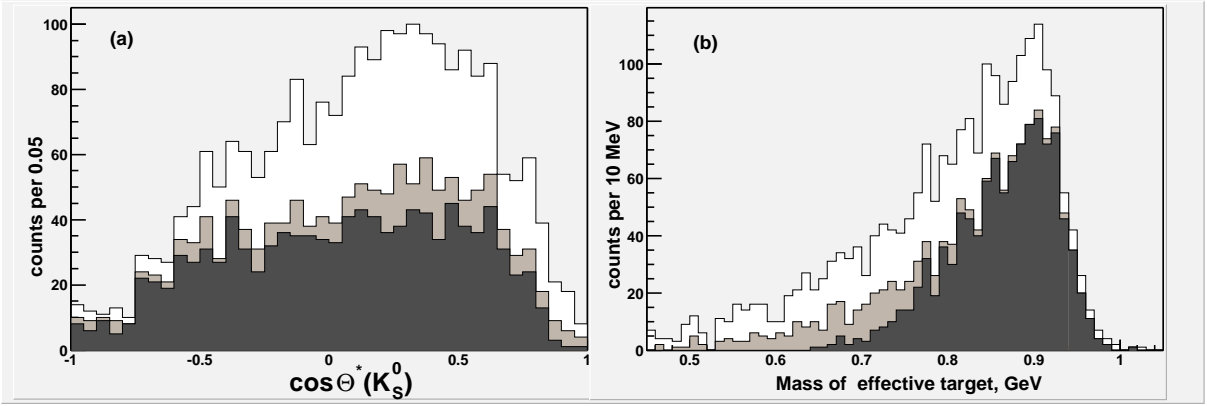}
\caption
{The cosine of the $K^0_S$ emission angle in the $pK^0_S$ frame (a)
and the effective target mass assuming binary kinematics (b). Shaded 
histograms are for the selection $|A_m| < 0.015$. Dark histograms 
result from an additional selection $p_L(pK^0_S) > 80$ MeV.}
\label{cosdec-and-mtar}
\end{figure}

     Qualitatively, $K^+$ collisions with bound and free neutrons
should form $pK^0_S$ systems with similar effective masses (prior to
proton and $K^0_S$ rescattering). The mass difference
$\Delta m = m(pK^0_S) - \sqrt{s}$ and the asymmetry
$A_m = ( m(pK^0_S) - \sqrt{s}) / ( m(pK^0_S) + \sqrt{s} )$
are plotted in Fig.~\ref{difma-and-asymma} for live and simulated
events of the charge-exchange reaction. For simulated events, the
``original" value of the $pK^0_S$ mass prior to any rescattering
is substituted.
The observed $\Delta m$ and $A_m$ distributions are seen to extend to 
negative values far in excess of the smearing expected from neutron 
binding in the Xe nucleus. In order to suppress rescattered events, 
we use the selection 
$|A_m| < 0.015$ 
suggested by the simulation.
The effect of this selection on the distribution of 
$\cos\Theta^*(K^0_S)$,
the cosine of the $K^0_S$ emission angle in the $pK^0_S$ rest frame, 
is shown in Fig.~\ref{cosdec-and-mtar}a.
By suppressing rescatterings, the selection $|A_m| < 0.015$ is seen
to render the latter distribution more uniform. 
On the other hand, this selection brings the \mtar\ distribution 
to better agreement with the mass of a bound neutron, see 
Fig.~\ref{cosdec-and-mtar}b. 

\begin{figure}[!t]

\vspace{10cm}
\includegraphics{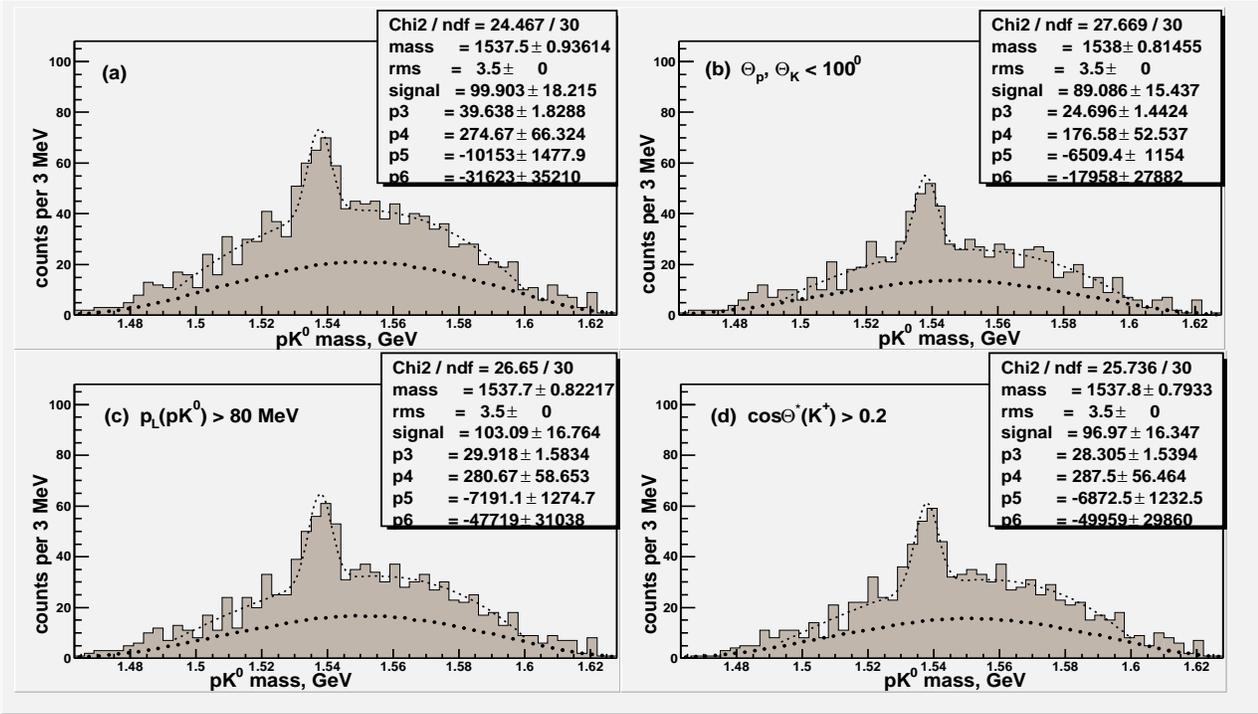}
\caption
{Effective mass of the $pK^0_S$ system upon applying the selection
$|A_m| < 0.015$ (a). An additional selection is
$\Theta_K < 100^0$ and $\Theta_p < 100^0$ in (b), 
$p_L(pK^0_S) > 80$ MeV in (c), and
$\cos\Theta^*(K^+) > 0.2$ in (d).
Depicted by dotted curves are fits 
to a Gaussian on top of a third-order polynomial. The width of the
Gaussian has been constrained to $\sigma = 3.5$ MeV.
The solid dots are simulated distributions assuming no rescatterings,
each normalized to the corresponding experimental distribution by
area outside the peak region $1529 < m(pK^0_S) < 1547$ MeV.
The simulated mass spectra have been scaled by a factor of 0.5 in 
order to avoid confusion with the fitting function.}
\label{cut-and-fitted}
\end{figure}
\begin{figure}[!t]

\vspace{10cm}
\includegraphics{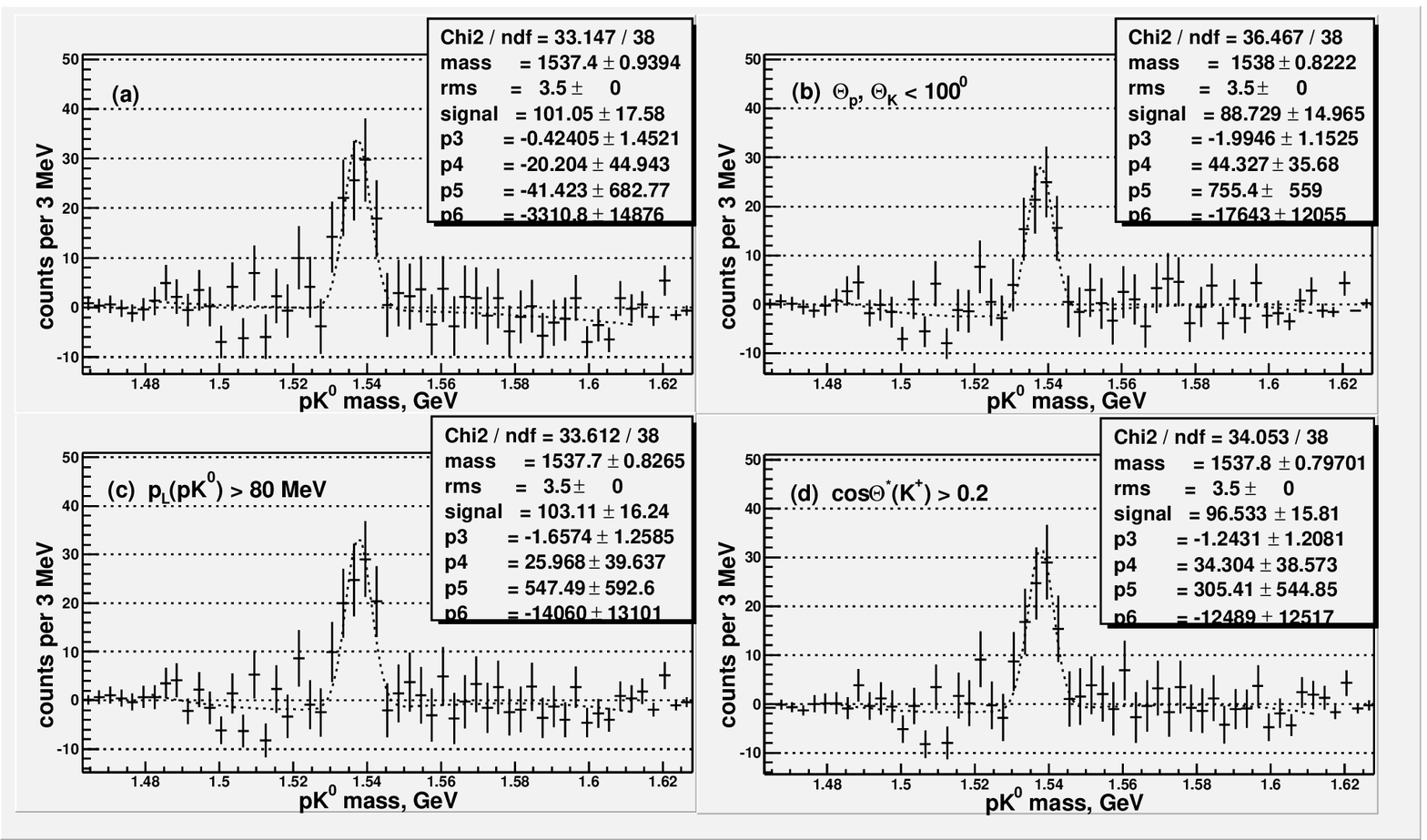}
\caption
{The difference between the experimental and simulated spectra of the
$pK^0_S$ effective mass, as shown in Fig.~\ref{cut-and-fitted}.}
\label{subtracted}
\end{figure}
     The \dimass\ distribution under the selection 
$|A_m| < 0.015$ shows an enhancement near 1538 MeV whose
width is consistent with instrumental smearing alone, see 
Fig.~\ref{cut-and-fitted}. In the fit of this \dimass\ spectrum 
to a Gaussian on top of a third-order polynomial, the width of the
Gaussian is constrained to $\sigma = 3.5$ MeV as estimated for a
zero-width resonance by propagating measurement errors for 
observed $pK^0_S$ pairs. This fit returns nearly 100 events above 
background. In order to further reduce the background from rescatterings,
an additinal selection is used: $\Theta_K < 100^0$ and 
$\Theta_p < 100^0$ for polar angles of the $K_S$ and proton as 
in \cite{DIANA}, or $p_L(pK^0_S) > 80$ MeV as suggested by 
Fig.~\ref{momenta}e, or
$\cos\Theta^*(K^+) > 0.2$
as suggested by Fig.~\ref{momenta}d.
Each additional selection further flattens the 
$\cos\Theta_\mathrm{cm}$ distribution and suppresses the downward
tail of the \mtar\ distribution (for $p_L > 80$ MeV, this is 
illustrated in Fig.~\ref{cosdec-and-mtar}). On the other hand, each 
selection further emphasizes the $pK^0_S$ peak near 1538 MeV, see 
Figs.~\ref{cut-and-fitted}b, \ref{cut-and-fitted}c, and
\ref{cut-and-fitted}d.
The statistical significance of the signal reaches some 8$\sigma$ 
(6$\sigma$) when estimated as $S /\sqrt{B}$ ($S /\sqrt{B+S}$). 
Also shown in  Fig.~\ref{cut-and-fitted} are the simulated distributions
of the $pK^0_S$ effective mass prior to any rescatterings under
similar selections. Each of these has been normalized to the
corresponding experimental distribution by area outside the peak region
$1529 < m(pK^0_S) < 1547$ MeV.
That peak position is very close to the maximum of the non-resonant
background may lead to overestimating the signal. So we additionally fit 
the difference between the experimental and simulated distributions,
see Fig.~\ref{subtracted}. The fits of subtracted mass spectra
return virtually the same signals as in Fig.~\ref{cut-and-fitted},
suggesting that the fitting procedure is robust.

\begin{figure}[h]

\vspace{5.5cm}
\includegraphics{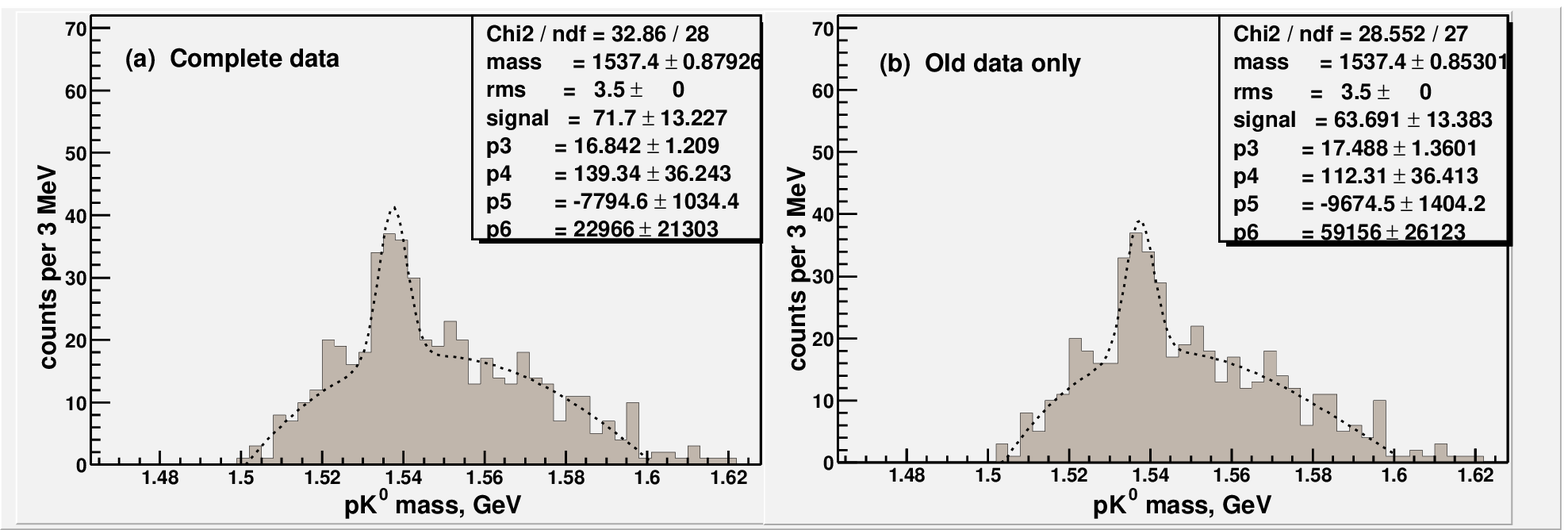}
\caption
{The $pK^0_S$ effective mass under the selections as in 
Fig.~\ref{cut-and-fitted}b 
($|A_m| < 0.015$, $\Theta_K < 100^0$, and $\Theta_p < 100^0$)
plus the additional selection  
$445 < p(K^+) < 525$ MeV. The data for the complete sample
of the charge-exchange reaction and for the part of the sample
analyzed in \cite{DIANA} are shown in (a) and (b), respectively.}
\label{window}
\end{figure}
     In ref.~\cite{DIANA} where the selections 
$\Theta_K < 100^0$ and $\Theta_p < 100^0$ 
were used, we argued that the bulk of the $\Theta^+$ signal should
arise from incident $K^+$ mesons with momenta in a restricted 
interval. Plotted in Fig.~\ref{window} is the $pK^0_S$ effective mass
under the selections as in Fig.~\ref{cut-and-fitted}b plus the
additional selection  
$445 < p(K^+) < 525$ MeV
as in \cite{DIANA}. This is separately done for the complete sample
of the charge-exchange reaction and for the part of the sample
analyzed in \cite{DIANA}. The signal in Fig.~\ref{window}b is 
virtually the same as that reported in \cite{DIANA}. Combining the
selections $|A_m| < 0.015$ and $445 < p(K^+) < 525$ MeV is of course
a redundant procedure used only for comparing
with our previous results. That the selection $|A_m| < 0.015$ is not
tuned to any particular value of the $pK^0_S$ effective mass is its
obvious advantage.

     Intrinsic width of a resonance formed in an $s$-channel reaction
like \under\ can be estimated by comparing the signal magnitude with
the level of non-resonant background under the peak, see {\it e.g.}
\cite{Cahn-Trilling}. 
However, this method cannot be directly applied
to $K^+$ collisions with heavy nuclei because of secondary interactions
of the $K^0$ and proton in nuclear medium. That is, the non-resonant
background under the peak is an unknown mixture of unrescattered and
rescattered events, whereas a true signal should consist of 
unrescattered events only. As soon as the $\Theta^+$ decay width is 
$\sim 1$ MeV or less, the peak will not be depleted by the $K^0_S$
and proton rescatterings because the bulk of produced $\Theta^+$ 
baryons will decay upon leaving the nucleus. The corresponding 
``rescattering-free" background can only be estimated by simulating
the effective mass of the originally produced $K^0_S$ and proton
prior to any rescatterings, 
$m_0(pK^0_S)$. 
\begin{figure}[h]

\vspace{6cm}
\includegraphics{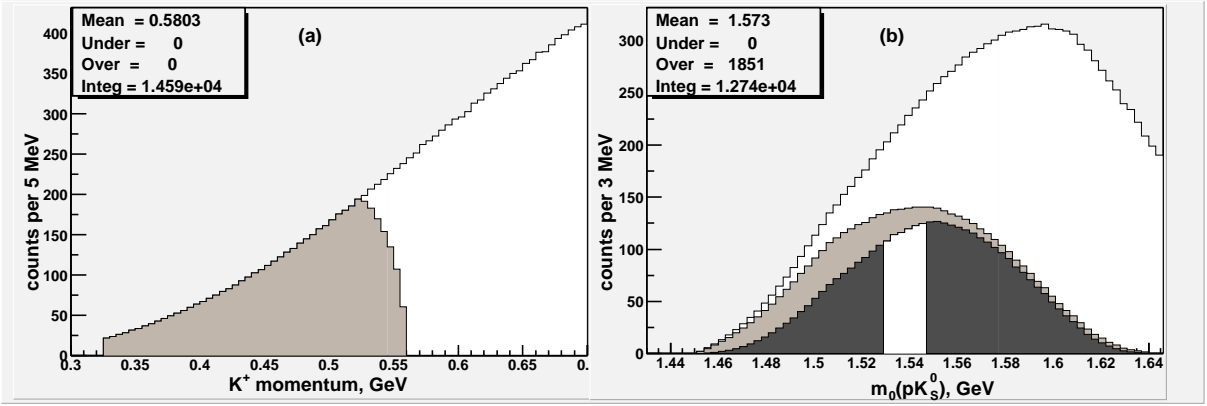}
\caption
{The $K^+$ momentum (a) and the ``undistorted" mass $m_0(pK^0_S)$ (b)
for simulated charge-exchange collisions \under\ in the bubble chamber
DIANA. The upper histograms are for all collisions 
with $p_K > 140$ MeV in the full fiducial
volume. The shaded histograms are for $K^+$ track lengths $L_K > 520$ mm 
(the region of throughput measurements). The dark-shaded histogram in 
(b) results from the adopted selections $p_K > 160$ MeV, $p_p > 170$ MeV, 
$\tau < 3\tau_0$,
and $|A_m| < 0.015$. The open-white
corridor in the latter histogram depicts the mass region
$1529 < m_0(pK^0_S) < 1547$ MeV. The normalization is 
explained in the text.}
\label{undistorted}
\end{figure}

     For the simulated charge-exchange collisions \under\ on a bound 
neutron in the bubble chamber DIANA, the $K^+$ momentum at collision 
point and the ``undistorted" mass $m_0(pK^0_S)$ are plotted in 
Fig.~\ref{undistorted}. The $K^0_S$ lab momentum is restricted to
$p_K > 140$ MeV which is the (effective) threshold for detecting
$K^0_S \rightarrow \pi^+\pi^-$ decays in the scan.
The upper histograms are for all collisions
in the full fiducial volume, {\it i.e.,} over the full range
of $L_K$---the $K^+$ track length in Xenon before the collision.
The shaded area of the $K^+$ momentum distribution corresponds 
to the restricted fiducial volume $L_K > 520$ mm where throughput
measurements were done. The absolute normalization of the simulated 
distributions of Fig.~\ref{undistorted} is based on the scanning
information. In the region $L_K > 520$ mm, the scan found $7900\pm500$
events with a $K^0_S \rightarrow \pi^+\pi^-$ candidate in the final 
state. Of these events, $(60\pm7)$\% are estimated to survive upon 
rejecting $K^0_S$ mesons with $L < 2.5$ mm, protons and $K^0_S$ mesons 
that have reinteracted in liquid Xenon, and unmeasurable events. The
resultant number of events, $4740\pm630$,
is then used for absolute
normalization of the simulated $K^+$-momentum distribution for 
$L_K > 520$ mm, see the shaded area in Fig.~\ref{undistorted}a.
The distribution of the same events in $m_0(pK^0_S)$ is shown by the 
corresponding shaded histogram in Fig.~\ref{undistorted}b. And finally, 
the absolute level of the non-resonant background can be estimated 
upon applying the aforementioned selections to simulated events: 
$p_K > 160$ MeV, $p_p > 170$ MeV, 
$\tau < 3\tau_0$,
and $|A_m| < 0.015$.
The resultant $m_0(pK^0_S)$ distribution, depicted by the 
dark-shaded histogram in Fig.~\ref{undistorted}b, features
$705\pm94$ events in the mass region 
$1529 < m_0(pK^0_S) < 1547$ MeV.

     Using the observed signal and simulated background as input and 
assuming $J = 1/2$, we then estimate the intrinsic width of the $pK^0$ 
resonance as \cite{Cahn-Trilling}
\begin{displaymath}
\Gamma = \frac{N^\mathrm{peak}}{N^\mathrm{bkgd}}
\times \frac{\sigma^\mathrm{CE}}{107\mathrm{mb}}
\times \frac{\Delta m}{B_i B_f},
\end{displaymath}
where $N^\mathrm{peak}$ and $N^\mathrm{bkgd}$ are the numbers of events
in the peak and in the background under the peak,
$\sigma^\mathrm{CE} = 4.1\pm0.3$ mb is the measured cross section  
of the charge-exchange reaction \under\ \cite{cross-section}, 
$B_i$ and $B_f$ are the branching fractions for the initial and final 
states ($B_i = B_f = 1/2$ for either $I = 0$ and $I = 1$), and 
$\Delta m$ is the \dimass\ interval under the peak that is populated by 
$N^\mathrm{bkgd}$ background events. Selecting a mass interval
$1529 < m(pK^0) < 1547$ MeV and substituting
$N^\mathrm{peak} = 99.9\pm18.2$ events from Fig.~\ref{cut-and-fitted}a
and $N^\mathrm{bkgd} = 705\pm94$
events from Fig.~\ref{undistorted}b,
we obtain $\Gamma = 0.39\pm0.10$ MeV where the error does not include
the systematic uncertainties of the simulation procedure. A systematic
shift of up to +20\% may result from taking into account the 
sequential reaction $K^+N \rightarrow K^+N$, \under\ \cite{Zorn}.
The value of $\Gamma$ derived in this analysis is consistent with our 
earlier estimate \cite{DIANA} and with the upper limit set by 
BELLE \cite{BELLE}.

     To conclude, we have reanalyzed the DIANA data on the 
charge-exchange reaction \charex\ using increased statistics and updated
selections. We have obtained new strong evidence for formation of a 
pentaquark baryon with positive strangeness in the charge-exchange 
reaction \under\ on a bound neutron. 
The statistical significance of the signal reaches some 8$\sigma$ 
(6$\sigma$) when estimated as $S /\sqrt{B}$ ($S /\sqrt{B+S}$). The
mass and intrinsic width of the $\Theta^+$ baryon are measured as
$m = 1538\pm2$ MeV and $\Gamma = 0.39\pm0.10$ MeV.

     We wish to thank Ya.~I. Azimov, L.~N. Bogdanova, and 
I.~I. Strakovsky for useful comments. This work is supported by 
the Russian Foundation for Basic Research (grant 07-02-00684).

\end{document}